\documentstyle[aps,preprint]{revtex}
\begin{document}
\draft
\tightenlines
%\preprint{gr-qc/9907031} \draft\tightenlines

\title{Gravitational Collapse and Calogero Model}
\author{Dongsu
Bak\footnote{Electronic address: dsbak@mach.uos.ac.kr}$^{a}$, Sang
Pyo Kim\footnote{Electronic address:
sangkim@knusun1.kunsan.ac.kr}$^{b}$, Sung Ku
Kim\footnote{Electronic address: skkim@theory.ewha.ac.kr}$^{c}$,
Kwang-Sup Soh\footnote{Electronic address:
kssoh@phya.snu.ac.kr}$^{d}$, and Jae Hyung Yee\footnote{Electronic
address: jhyee@phya.yonsei.ac.kr}$^{e}$}
\address{a Department of Physics, University of Seoul, Seoul
130-743 Korea\\ b Department of Physics, Kunsan National
University, Kunsan 573-701 Korea\\ c Department of Physics, Ewha
Women's University, Seoul 120-750 Korea\\ d Department of Physics
Education, Seoul National University, Seoul 151-742 Korea\\ e
Department of Physics and Institute of Physics and Applied Physics, 
Yonsei University, Seoul 120-749 Korea}

\date{\today}

\maketitle

\begin{abstract}
We study the analytic structure of the S-matrix which is obtained from
the reduced Wheeler-DeWitt wave function describing spherically
symmetric gravitational collapse of massless scalar fields.  The simple
poles in the S-matrix occur in the Euclidean spacetime, and the
Euclidean Wheeler-DeWitt equation is a variant of the Calogero models,
which is
discussed in connection with conformal mechanics and a quantum
instanton.
\end{abstract}

In the previous work \cite{pre} we studied quantum mechanically the
self-similar black hole formation by collapsing scalar fields and found
the wave functions that give the correct semi-classical limit.  In this
brief report we consider the pole structure of the S-matrix which is
obtained from the wave function.  The pole corresponds to a solution in
an unphysical region, namely, in the Euclidean spacetime with a
quantized parameter($c_0$), which seems like a quantum version of
instanton.

The spherically symmetric geometry minimally coupled to a massless
scalar field is described by the reduced action in $(1+1)$-dimensional
spacetime of which the Hilbert-Einstein action is

\begin{equation}
S = \frac{1}{16\pi} \int_M d^4x \sqrt{-g}~ \left[ R - 2 \left(\nabla \phi
\right) ^2 \right] + \frac{1}{8\pi} \int_{\partial M} d^3 x K \sqrt{h}.
\end{equation}
The reduced action is
\begin{equation}
S_{sph} = \frac{1}{4} \int d^2 x \sqrt{-\gamma} ~r^2 \left[ \left\{^{(2)}
R(\gamma) + \frac{2}{r^2} \left(\left(\nabla r \right)^2 + 1\right)
\right\} -2 \left(\nabla\phi\right)^2 \right],
\end{equation}
where $\gamma_{ab}$ is the $(1+1)$-dimensional metric.  The spherical
spacetime
metric is
\begin{equation}
ds^2 = -2 du~dv + r^2 d\Omega_2^2 ,
\end{equation}
where $d\Omega_2^2$ is the usual spherical part of the metric, and $u$
and $v$ are null coordinates.  The self-similarity condition is imposed
such that 
\begin{equation}
r = \sqrt{-uv}~ y(z), \hspace{3mm} \phi = \phi(z),
\end{equation}
where $z = \frac{+v}{-u} = e^{-2\tau}$, $y$ and $\phi$ depend only on $z$.
We introduce another coordinates $(\omega, \tau)$ as
\begin{equation}
u = -\omega e^{-\tau}, \hspace{3mm} v = \omega e^\tau ,
\end{equation}

\begin{equation}
ds^2 = -2 \omega^2 d\tau^2 + 2d\omega^2 + \omega^2 y^2 d\Omega_2^2.
\end{equation}

The classical solutions of the field equations were obtained by Roberts
\cite{Roberts}, and studied in connection with gravitational collapse by
others \cite{others}.  Classically black hole formation was only allowed
in the supercritical cases ($c_0 >1$), but even in the subcritical
situation there are quantum mechanical tunneling processes to form a
black hole of which the probability is semiclassically calculated
\cite{subcritical}.

In our previous work \cite{pre} we quantized the system canonically with
the ADM formulation to obtain the Wheeler-DeWitt equation for the
quantum black hole formation
\begin{equation}
\left[\frac{1}{2K} \frac{\partial^2}{\partial y^2} - \frac{1}{2Ky^2}
\frac{\partial^2}{\partial\phi^2} - K\left(1-\frac{y^2}{2}\right)
\right] \Psi(y,\phi) = 0,
\end{equation}
where $K\equiv \frac{m_p^2}{\hbar} \frac{\omega_c^2}{2}$ plays the role of
a cut-off parameter of the model, and we use a unit system $\hbar =1$,
$m_p=1$, and $c=1$.  The wave function can be factorized to the scalar
and gravitational parts,
\begin{equation}
\Psi(y,\phi) = \exp\left(\pm i K c_0 \phi\right) \psi(y).
\end{equation}
Here the scalar field part is chosen to yield the classical momentum
$\pi_\phi = \pm Kc_0$, where $c_0$ is the dimensionless parameter to
determining the supercritical ($c_0 > 1$), the critical ($c_0 =1$), and
the
subcritical($1>c_0 >0$) collapse. 

Now the Wheeler-DeWitt equation becomes an ordinary differential
equation
\begin{equation}
\left[\frac{-1}{2K} \frac{d^2}{dy^2} + \frac{K}{2}\left(2 - y^2 -
\frac{c_0^2}{y^2} \right) \right] \psi(y) =0.
\end{equation}
The solution describing the black hole formation was obtained in the
Ref.\cite{pre}:
\begin{equation}
\psi_{BH} (y) = \left(\exp\frac{-i}{2} K y^2 \right) \left(K y^2
\right)^{\mu_-} M\left( a_-, b_-, iKy^2\right),
\end{equation}
of which the asymptotic form \cite{asymptotic} at the spatial infinity
is 
\begin{equation}
\psi_{BH} (y) \simeq \Gamma (b_-) \left[ \frac{e^{i\pi
a_-}}{\Gamma(a_+^*)} \left(iKy^2 \right)^{\mu_- - a_-} e^{-\frac{i}{2} K
y^2} + \frac{\left(iKy^2 \right)^{\mu_- - a_+^*}}{\Gamma(a_-)}
e^{\frac{i}{2} Ky^2} \right].
\end{equation}
Here $M$ is the confluent hypergeometric function and 
\begin{equation}
a_\pm = \frac{1}{2} \pm \frac{i}{2} (Q\mp K), \hspace{2mm} b_- = 1-iQ,
\hspace{2mm} \mu_- = \frac{1}{4} - \frac{i}{2}Q
\end{equation}
with
\begin{equation}
Q=\left( K^2 c_0^2 -\frac{1}{4}\right)^{1/2}.
\end{equation}
The S-matrix component describing the reflection rate is 
\begin{equation}
S = \frac{\Gamma (a_+^*)}{\Gamma(a_-)} \frac{(iK)^{a_- - a_+^*}}{e^{i\pi
a_-}},
\end{equation}
from which we obtain the transmission rate for black hole formation
\begin{eqnarray}
\label{trans}
\frac{j_{trans}}{j_{in}} & = & 1 - |S|^2 \nonumber \\
& = & 1- \frac{\cosh \frac{\pi}{2}(Q+K)}{\cosh \frac{\pi}{2}(Q-K)} e^{-\pi
Q} , 
\end{eqnarray}
where $\left|\Gamma \left(\frac{1}{2}+ix\right)\right|^2 = \pi /
\cosh(\pi x)$ is used.  Eq. (\ref{trans}) gives the probability of
black hole formation for the supercritical, critical, and subcritical
$c_0$-values.

In this brief report we consider the analytic structure of the S-matrix
: It is an analytic function of $Q$ and $K$ with simple poles which can
be explicitly shown as
\begin{equation}
\label{Smatrix}
S = \sum_{N=0}^{\infty} \frac{1}{Q-K+i(2N+1)} \left( \frac{2i
e^{-\frac{\pi}{2}K-iK \ln K}}{N! \Gamma(-N-iK)}\right).
\end{equation}
The poles reside in the unphysical region of the parameter space of $Q$
and $K$:
\begin{equation}
\label{Q}
Q = \sqrt{K^2 c_0^2 - \frac{1}{4}} = K - i(2N+1) , \hspace{4mm} N =
0,1,2, \cdots .
\end{equation}

For physical processes of gravitational collapse there can not be poles
because $K$ and $c_0$ are real valued.
In ordinary quantum mechanics, the poles of S-matrix occur at the bound
state \cite{poles}, and in relativistic scattering at the resonances or
the Regge poles \cite{Regge}.  In our case we can not identify the poles
with such bound
states or resonances.

For physical interpretation of the poles we note that the replacement 
\begin{equation}
K = i K_E
\end{equation}
is, in effect, the change from the Lorentzian metric to the Euclidean
one, and the Wheeler-DeWitt equation in the Euclidean sector becomes 
\begin{equation}
\label{WDeq}
\left[ \frac{-1}{2K_E} \frac{d^2}{dy^2} + \frac{K_E}{2} \left( y^2 +
\frac{c_0^2}{y^2} -2\right) \right] \psi_E =0.
\end{equation}
Notice that this is a variant of Calogero models with the Calogero-Moser
hamiltonian \cite{Calogero}, but the energy eigenvalue is fixed, and only
a quantized $c_0$ is allowed.  The solution to the equation is of
polynomial type
\begin{equation}
\label{Quantum}
\psi_E = e^{-\frac{1}{2}K_E y^2} \left( y \sqrt{2K_E}\right)^{K_E - 2N
-\frac{1}{2}} \left( \sum_{m=0}^{N} a_m
\left(\sqrt{2K_E}y\right)^m\right),
\end{equation}
\begin{equation}
a_{m+2} = \frac{m-2N}{(m+2)(m+2K_E -4N)}a_m.
\end{equation}
Here we use the normalizability condition of the wave function $\psi_E$,
which gives the quantization of $c_0$ as
\begin{equation}
\label{nzcondition}
K_E^2 c_0^2 + \frac{1}{4} = \left( K_E -2N -1\right)^2 \hspace{3mm},
\end{equation}
or
\begin{equation}
c_0^2 = \left(1-\frac{2N-1}{K_E}\right)^2 -\frac{1}{4K_E^2}.
\end{equation}
The condition (\ref{nzcondition}) is identical to the pole position of
the S-matrix with $K=iK_E$ given in (\ref{Q}).

The quantum solution (\ref{Quantum}) is analogous to an instanton in the
sense it is a solution in the Euclidean sector.
However, it is not an instanton which is strictly a classical solution.
With instantons one can semiclassically evaluate the probability of
tunneling process, while our solutions (\ref{Quantum}) provide the quantum
probability through the pole contribution to the matrix as in
(\ref{Smatrix}).

The correspondence between the poles and the Euclidean polynomial
solutions breaks down for large $N$.  While the poles contribute for all
$N$ without limit, the normalizable Euclidean solutions exist only for
$N < \frac{K_E}{2}$.  The polynomial solutions for large $N$ are well
defined, but are not normalizable.  We have not yet understood these
nonnormalizable solutions.

One notable point of the Euclidean Wheeler-DeWitt equation is that it is
a Calogero type system \cite{Calogero} which has been extensively
discussed in connection with the motion of a particle near the
Reissner-Nordstr\"{o}m black hole horizon \cite{RN}.  A surprising
connection
between black holes and conformal mechanics of De Alfaro, Fubini and
Furlan \cite{DFF} (DFF) has been established \cite{connection}.  
By studying the conformal quantum mechanics DFF suggested to solve the
problem of motion to use a compact operator $L_0$ given as
\begin{equation}
L_0 = \frac{a}{4} \left( \frac{x^2}{a^2} + p^2 + \frac{g}{x^2} \right),
\end{equation}
which is essentially identical to the Wheeler-DeWitt hamiltonian
(\ref{WDeq}).  The presence of conformal mechanics both in the
Reissner-Nordstr\"{o}m black hole and in the gravitational collapse
forming
black holes might suggest some deep nature of gravity in connection with
conformal theory.

As a final remark we consider the classical field equations
corresponding to the poles of the S-matrix.  The relevant equations are
\begin{equation}
\frac{d\phi}{d\tau} = \frac{c_0}{y^2}, 
\end{equation}
\begin{equation}
\left( \frac{dy}{d\tau} \right)^2 = K^2 \left( -2 + y^2 +
\frac{c_0^2}{y^2} \right),
\end{equation}
where in the Lorentz metric spacetime $c_0 \simeq 1- i (2N+1) / K$, for
large $K$.  The complex $c_0$ implies complex $\frac{d\phi}{d\tau}$,
which may be imagined as a bound state like complex momentum in quantum
mechanics.  The complex $\frac{dy}{d\tau}$ is difficult to understand
unless one considers complex spacetime metric.  
In the Euclidean case ($K = iK_E$) these classical equations are same as
those equations with quantized $c_0$ in the tunneling region in Ref.
\cite{subcritical}.  It is beyond the scope of our present
work to investigate the role of complex spacetime in gravitational
collapse.

\begin{acknowledgements}
This work was supported in parts by BSRI Program under BSRI
98-015-D00054, 98-015-D00061, 98-015-D00129, and by KOSEF through the
CTP of Seoul National University.
\end{acknowledgements}

\end{document}